\documentclass{elsart1p}

 \usepackage{graphicx}

\usepackage{amssymb}

\begin{document}

\begin{frontmatter}



\title{Can a strong magnetic background modify the nature of the chiral transition in QCD?}


\author{Eduardo S. Fraga and Ana J\'ulia Mizher}

\address{Instituto de F\'\i sica, Universidade Federal do Rio de Janeiro \\
Caixa Postal 68528, Rio de Janeiro, RJ 21941-972, Brazil}

\begin{abstract}
The presence of a strong magnetic background can modify the nature and the 
dynamics of the chiral phase transition at finite temperature: for high enough 
magnetic fields, comparable to the ones expected to be created in noncentral 
high-energy heavy ion collisions at RHIC and the LHC, the original crossover is 
turned into a first-order transition. We illustrate this effect within the linear sigma 
model with quarks to one loop in the $\overline{\rm MS}$ scheme for $N_{f}=2$. 
\end{abstract}

\end{frontmatter}


Strong magnetic fields are known to produce remarkable physical effects in macroscopic 
compact astrophysical objects, such as illustrated by magnetars \cite{magnetars}. 
More recently, it has been proposed that very high magnetic fields could be created 
in noncentral high-energy heavy ion collisions at the Relativistic Heavy Ion Collider (RHIC), 
at Brookhaven, and the Large Hadron Collider (LHC), at CERN, affecting observables 
that could provide a signature of the presence of CP-odd domains in the presumably formed 
quark-gluon plasma (QGP) phase \cite{CP-odd} via a mechanism of separation of 
charge. Estimates point to magnetic fields $B\sim 10^{19}~$ Gauss ($eB\sim 6 m_{\pi}^{2}$), 
a very intense magnetic background from the point of view of quantum chromodynamics (QCD). 

Besides its role in the separation of charge in CP-odd domains, one can ask a simpler 
and no less interesting question about the possible effects of high magnetic fields 
on hot quark matter \cite{Fraga:2008qn}: can a strong magnetic background modify the 
nature of the chiral transition in QCD? We show below that the answer is yes, and that, for high 
enough magnetic fields, comparable to the ones expected to be created in noncentral 
high-energy heavy ion collisions at RHIC and the LHC, the original crossover, as predicted 
by lattice simulations, is turned into a first-order transition \cite{Fraga:2008qn}. 

Modifications in the vacuum of quantum electrodynamics and QCD have also been 
investigated within different frameworks, mainly using effective 
models \cite{Klevansky:1989vi,Gusynin:1994xp,Babansky:1997zh,Klimenko:1998su,
Semenoff:1999xv,Goyal:1999ye,Hiller:2008eh,Rojas:2008sg}, 
especially the NJL model, and chiral perturbation 
theory \cite{Shushpanov:1997sf,Agasian:1999sx,Cohen:2007bt}, but also resorting to the 
quark model \cite{Kabat:2002er} and certain limits of QCD \cite{Miransky:2002rp}. 
Most treatments have been concerned with vacuum modifications by the magnetic field, 
though medium effects were considered in a few cases. More recently, effects on the dynamical 
quark mass \cite{Klimenko:2008mg} and on the quark-hadron transition \cite{Agasian:2008tb} 
were also considered. 

To investigate the effects of a strong magnetic background on the nature and 
dynamics of the chiral phase transition at finite temperature, $T$, and vanishing chemical 
potential, we adopt as an effective theory the linear sigma model coupled to quarks (LSM$_{q}$) 
with two flavors, $N_{f}=2$ \cite{GellMann:1960np,Scavenius:2000qd}, defined by the lagrangian
\begin{eqnarray}
{\cal L} &=&
 \overline{\psi}_f \left[i\gamma ^{\mu}\partial _{\mu} - g(\sigma +i\gamma _{5}
 \vec{\tau} \cdot \vec{\pi} )\right]\psi_f 
+ \frac{1}{2}(\partial _{\mu}\sigma \partial ^{\mu}\sigma + \partial _{\mu}
\vec{\pi} \partial ^{\mu}\vec{\pi} )
- V(\sigma ,\vec{\pi})\;,
\label{lagrangian}
\end{eqnarray}
where $V(\sigma ,\vec{\pi})=\frac{\lambda}{4}(\sigma^{2}+\vec{\pi}^{2} -
{\it v}^2)^2-h\sigma$ 
is the self-interaction potential for the mesons, exhibiting both spontaneous 
and explicit breaking of chiral symmetry. The $N_f=2$ massive fermion fields 
$\psi_f$ represent the up and down constituent-quark fields $\psi=(u,d)$. The 
scalar field $\sigma$ plays the role of an approximate order parameter for the 
chiral transition, being an exact order parameter for massless quarks and pions. 
The latter are represented by the pseudoscalar field $\vec{\pi}=(\pi^{0},\pi^{+},\pi^{-})$, 
and it is common to group together these meson fields into an $O(4)$ chiral field 
$\phi =(\sigma,\vec{\pi})$. 
In what follows, we implement a simple mean-field treatment with the customary simplifying 
assumptions (see, e.g., Ref. \cite{Scavenius:2000qd}). Quarks constitute a thermalized fluid 
that provides a background in which the long wavelength modes of the chiral condensate evolve. 
Hence, at $T=0$, the model reproduces results from the usual LSM without 
quarks or from chiral perturbation theory for the broken phase vacuum. 
In this phase, quark degrees of freedom are absent (excited only for $T > 0$). The $\sigma$ 
field is heavy, $M_{\sigma} \sim600$ MeV, and treated classically. On the other hand, pions are 
light, and fluctuations in $\pi^{+}$  and $\pi^{-}$ couple to the magnetic field, $B$, as will be 
discussed in the next section, whereas fluctuations in $\pi^{0}$ give a $B$-independent 
contribution that we ignore, for simplicity. For $T > 0$, quarks are relevant (fast) degrees of 
freedom and chiral symmetry is approximately restored in the plasma for high enough $T$. 
In this case, we incorporate quark thermal fluctuations in the effective potential for $\sigma$, 
i.e. we integrate over quarks to one loop. Pions become rapidly heavy only after $T_{c}$ and 
their fluctuations can, in principle, matter since they couple to $B$. 
The parameters of the lagrangian are chosen such that the effective model reproduces 
correctly the phenomenology of QCD at low energies and in the vacuum, in the absence of 
a magnetic field.
Standard integration over the fermionic degrees of freedom to one loop,  
using a classical approximation for the chiral field, gives the effective potential 
in the $\sigma$ direction $V_{eff}= V(\phi)+V_q(\phi)$, where $V_{q}$ represents the 
thermal contribution from the quarks that acquire an effective mass $M(\sigma)=g|\sigma|$.
The net effect of the term $V_{q}$ is correcting the potential for the chiral 
field, approximately restoring chiral symmetry for a critical temperature 
$T_{c}\sim 150~$MeV \cite{Scavenius:2000qd}. 


Assuming that the system is now in the presence of a strong magnetic field background that is constant 
and homogeneous, one can compute the modified effective potential following the procedure outlined 
in Ref. \cite{Fraga:2008qn}. In what follows, we simply provide some of the main results. 
For definiteness, let us take the direction of the magnetic field as the $z$-direction, $\vec{B}=B \hat z$. 
The effective potential can be generalized to this case by a simple redefinition of the dispersion relations 
of the fields in the presence of $\vec{B}$, using the minimal coupling shift in the gradient and the field 
equations of motion. For this purpose, it is convenient to choose the gauge such that 
$A^{\mu}=(A^{0},\vec{A})=(0,-By,0,0)$. Decomposing the fields into their Fourier modes, one arrives 
at eigenvalue equations which have the same form as the Schr\"odinger equation for a harmonic 
oscillator potential, whose eigenmodes correspond to the well-known Landau levels. The latter 
provide the new dispersion relations
\begin{eqnarray}
p_{0n}^2=p_z^2+m^2+(2n+1)|q|B \quad , \quad
p_{0n}^2=p_z^2+m^2+(2n+1-\sigma)|q|B \, ,
\end{eqnarray}
for scalars and fermions, respectively, $n$ being an integer, $q$ the electric charge, 
and $\sigma$ the sign of the spin. Integrals over four momenta and thermal 
sum-integrals are modified accordingly, yielding sums over the Landau levels.

In our effective model, the vacuum piece of the potential will be modified by the magnetic field 
through the coupling of the field to charged pions. To one loop, and in the limit 
of high $B$, $eB >> m_{\pi}^{2}$, one obtains (ignoring contributions independent of the 
condensates) \cite{Fraga:2008qn}
\begin{equation}
V_{\pi^+}^V+V_{\pi^-}^V=-\frac{2m_\pi^2 eB}{32\pi^2}\log 2 \, .
\label{Vpion}
\end{equation}

Thermal corrections are provided by pions and quarks. However, the pion thermal contribution 
as well as part of the quark thermal contribution are exponentially suppressed for high magnetic 
fields, as has been shown in Ref. \cite{Fraga:2008qn}. The only part of the quark thermal piece 
that contributes is
\begin{equation}
V_q^T = -N_c \frac{eBT^2}{2\pi^2} \left[\int_{-\infty}^{+\infty} dx~ 
\ln\left( 1+e^{-\sqrt{x^2 +M_q^2/T^2}}\right)\right] \; ,
\label{Vquark}
\end{equation}
where $N_{c}=3$ is the number of colors. Therefore, the effective potential is corrected by the 
contributions in (\ref{Vpion}) and (\ref{Vquark}) in the presence of a strong homogeneous magnetic 
background. Therefore, the presence of the magnetic field enhances the value of the chiral condensate 
and the depth of the broken phase minimum of the modified effective potential, a result that is in line 
with those found within different approaches (see, for instance, 
Refs. \cite{Hiller:2008eh,Shushpanov:1997sf,Cohen:2007bt}).


At RHIC, estimates by Kharzeev, McLerran and Warringa \cite{CP-odd} give $eB\sim 5.3 m_{\pi}^{2}$. 
(For the LHC, there is a small increase). So, we adopt $eB\sim 6 m_{\pi}^{2}$ as a reasonable 
estimate. Fig. 1 displays the effective potential for $eB\sim 10 m_{\pi}^{2}$ at different values of the 
temperature to illustrate the phenomenon of chiral symmetry restoration via a first-order 
transition. For lower values of the field, the barrier is smaller. 
In Fig. 2, we show a zoom of the effective potential for $eB\sim 6 m_{\pi}^{2}$ for a temperature 
slightly below the critical one. This figure highlights the presence of a first-order barrier in 
the effective potential. For a magnetic field of the magnitude that could 
possibly be found in non-central high-energy heavy ion collisions, one moves from a crossover 
scenario to that of a weak first-order chiral transition, with a critical temperature 
$\sim 30\%$ higher \cite{Fraga:2008qn}.

\begin{figure}[htb]
\vspace{0.5cm}
\begin{minipage}[t]{56mm}
\includegraphics[width=6.1cm]{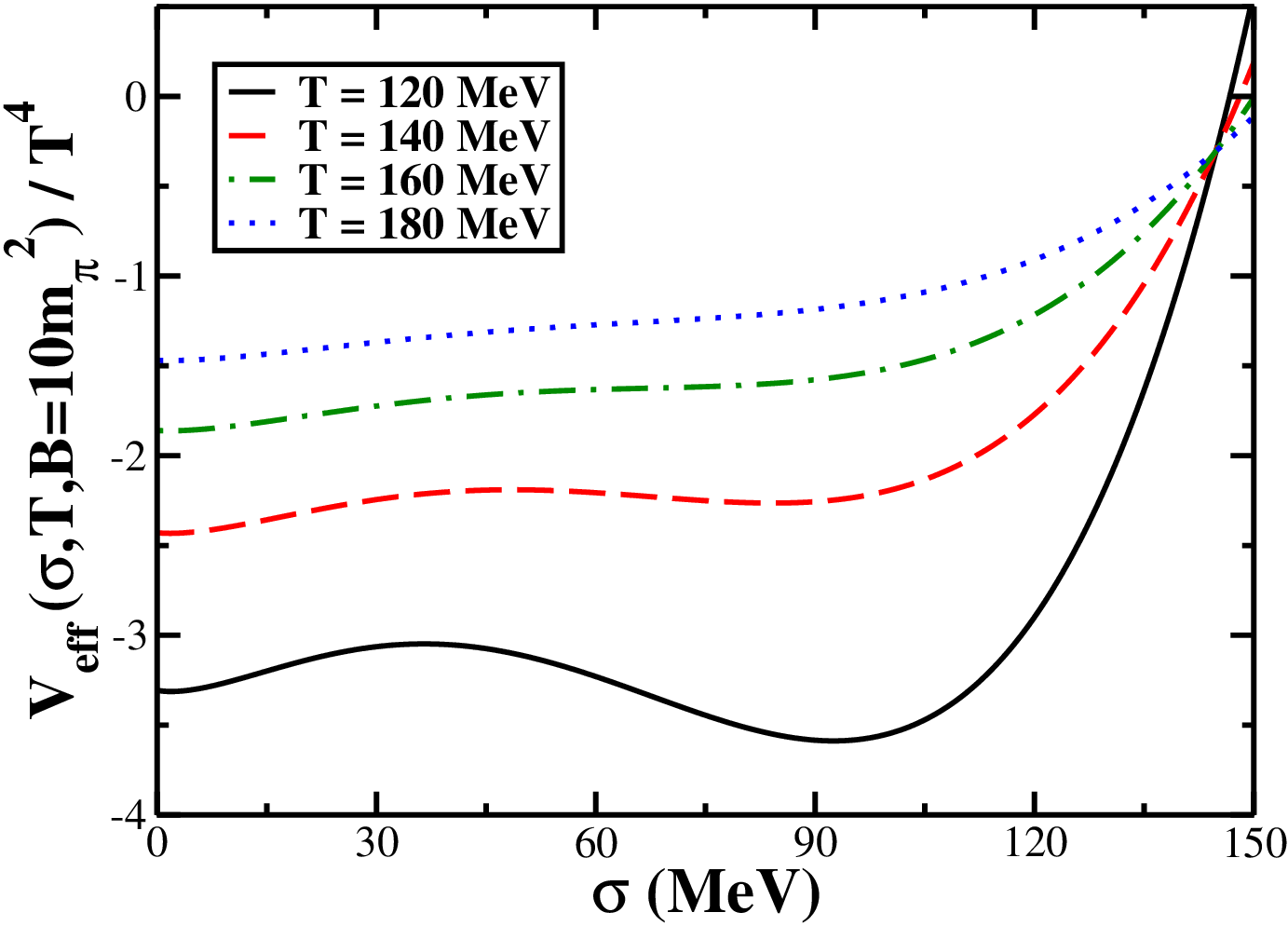}
\caption{Effective potential for $eB=10 m_{\pi}^{2}$.}
\label{V}
\end{minipage}
\hspace{1cm}
\begin{minipage}[t]{56mm}
\includegraphics[width=6.3cm]{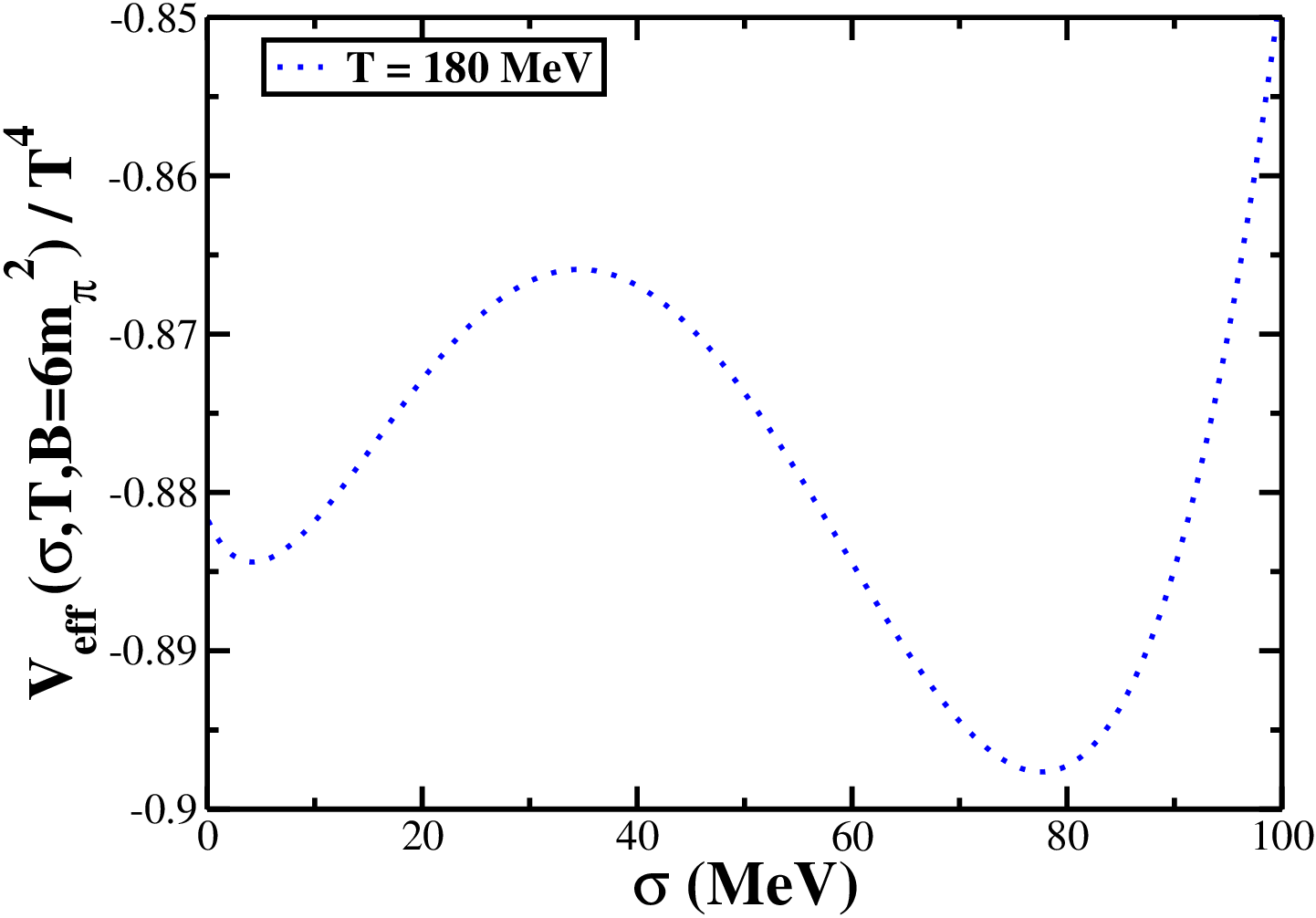}
\caption{Zoom of the barrier for $eB=6 m_{\pi}^{2}$.}
\label{Veff_B_6_zoom}
\end{minipage}
\end{figure}
%


Although Lattice QCD seems to indicate a crossover instead of a first-order chiral transition at 
vanishing chemical potential, a strong magnetic background might modify this picture dramatically. 
In particular, the results presented here might be relevant for the physics of the primordial QCD 
transition, where the description of phase conversion is generally via bubble nucleation, assuming 
a first-order transition \cite{Schwarz:2003du}. For heavy ion collisions at RHIC and LHC, although 
the barrier in the effective potential seems to be quite small, it can probably trap most of the system 
down to the spinodal explosion, altering the dynamics of phase conversion and its relevant 
time scales \cite{Fraga:2004hp}. This opens a new direction in the study of the phase diagram 
for strong interactions, with a rich phenomenology awaiting to be investigated \cite{Fraga:2008qn}. 
This work was partially supported by CAPES, CNPq, FAPERJ and FUJB/UFRJ.




\end{document}